\documentstyle[aps,graphicx,epsf]{revtex}\tighten
\def \BE {\begin{equation}}
\def \EE {\end{equation}}
\def \nt {\tau}

\def \pul {{{\scriptstyle{\frac{1}{2}}}}}

\def \ol {\overline}

\begin{document}

\title{A Class of Exact Classical Solutions to String Theory}
\author{A. A. Coley  } 
\address{Department of Mathematics and Statistics,
Dalhousie University, Halifax, Nova Scotia}
\maketitle
\begin{abstract}

We show that the
recently obtained class of spacetimes for
which all of the~scalar curvature invariants  
vanish (which can be regarded as generalizations of pp-wave spacetimes)
are exact solutions in string theory
to all perturbative orders in the~string tension scale.
As a result the~spectrum of
the~theory can be explicitly obtained, and 
these spacetimes are expected to provide some hints for
the~study of superstrings on more general backgrounds.
Since these Lorentzian  spacetimes suffer no quantum corrections
to all loop orders they may also offer insights into  quantum gravity.

\end{abstract}

\section{Introduction}

It is known that all of the~scalar curvature invariants  
vanish in pp-wave spacetimes \cite{jordan} (see also \cite{deser,gibb}).
It has subsequently been argued that
pp-wave spacetimes are exact solutions in string theory
(to all perturbative orders in the~string tension) 
\cite{amati,HS}.  
In this Letter we shall show that this is true for
a wide class of spacetimes (in addition to
the pp-wave spacetimes), a result that has 
broad and important implications.

In a recent  paper it was proven that in pseudo-Riemannian 
or Lorentzian spacetimes
all of the~scalar invariants
constructed from the~Riemann tensor and its 
covariant derivatives are
zero if and only if the~spacetime is of Petrov type 
III, N or O, all eigenvalues of
the~Ricci tensor are zero and the~common multiple null 
eigenvector of the~Weyl and Ricci tensors
is shearfree, irrotational, geodesic and expansion-free (SIGE)
\cite{cppm}; we shall refer to
these spacetimes as vanishing scalar invariant (VSI) 
spacetimes in what follows for brevity.
Utilizing a complex null tetrad in the Newman-Penrose 
(NP) formalism it was shown that
for Petrov types III and N the~repeated null vector of the~Weyl tensor 
$l^\alpha$ is SIGE
(i.e., the NP coefficients 
$\kappa$, $\sigma$, and $\rho$ are zero),
 and the~Ricci tensor has the~form  
\begin{equation}
 R_{\alpha \beta} = - 2\Phi_{22} l_{\alpha} 
 l_{\beta} + 4 \Phi_{21}  l_{(\alpha} m_{\beta)} 
+ 4 \Phi_{12}  l_{(\alpha} {\bar m}_{\beta)}, \label{Ricci}
\end{equation}
in terms of the non-zero Ricci components  
$\Phi_{ij}$.  
For Petrov type 0, the~Weyl tensor vanishes and so it 
suffices that the~Ricci tensor 
has the~form (\ref{Ricci}), where the~corresponding vector 
field $l^\alpha$ is again SIGE.

All of these spacetimes belong to Kundt's class, and hence
the~metric of these spacetimes can be
expressed  \cite{kramer,kundt}
\BE
{\rm d}s^2=2{\rm d} u [ H {\rm d}u+{\rm d} v
+ W  {\rm d}\zeta
+{\bar W} 
 {\rm d}{\bar \zeta}]
-2 {\rm d}\zeta{\rm d}{\bar\zeta}\ ,\label{dsKundt}
\EE
where $H=H(u,v,\zeta,{\bar\zeta})$ and $W= W(u,v,\zeta,{\bar\zeta})$ ($P \equiv 1$), 
and the~null tetrad is 
\BE
l=\partial_v, ~n=\partial_u-(H+W{\bar W})\partial_v+({\bar W}
\partial_\zeta+W\partial_{\bar\zeta}),
~m=\partial_\zeta\ .\label{Kundttetrada}
\EE
Note that in local coordinates the  repeated null Weyl eigenvector is given by $l=\partial_v$.
If  $\nt=0$, the null congruence $l$ is recurrent \cite{cppm}.

The  metrics  for 
all VSI spacetimes are displayed in  \cite{cppm}. 
For example, the metric of Pleba\` nski-Petrov  (PP)-type
O, Petrov type III and NP coefficient  $\nt=0$ 
is given by
\BE
W=W_0(u,{\bar\zeta}), ~
H=\pul v(W_0,_{\bar\zeta}+{\bar W}_0,_\zeta)
+h_0(u,\zeta,{\bar\zeta}),\label{example}
\EE
where
\BE
\Phi_{22}
=[h_0,_{\zeta{\bar\zeta}}-\Re (W_0W_0,_{{\bar\zeta}{\bar\zeta}}
+W_0,_{u{\bar\zeta}}+{W_0,_{\bar\zeta}}^2)],\label{example2}
\EE
and $l$ is recurrent.
The generalized pp-wave 
solutions are of Petrov-type N, PP-type O
(so that the~Ricci tensor has the~form of null radiation)
with $\nt=0$, and admit a covariantly constant null vector field 
\cite{jordan}. The
vacuum spacetimes, which are obtained by setting $\Phi_{22}= 0$,
are the~well-known
pp-wave spacetimes (or plane-fronted gravitational
waves with parallel rays).

The~Ricci tensor (\ref{Ricci})  has four vanishing
eigenvalues, and the PP-type is N    for
$\Phi_{12}\not= 0$
or O for
$\Phi_{12}=0$. 
It is known that the energy conditions are violated 
in the 
PP-type N models \cite{kramer} and hence attention is usually
concentrated on the
more physically interesting  PP-type O case, which in the  non-vacuum
case corresponds to 
a pure radiation  \cite{joly,kramer} 
(although we should point out that
spacetimes that violate the energy conditions are 
also of interest in current
applications \cite{energy}).

The~pp-wave spacetimes have a number of important physical
applications. In particular,
pp-wave spacetimes are exact solutions in string theory
(to all perturbative orders in the~string tension) 
\cite{amati,HS}.  We shall show 
that a wide range of VSI~spacetimes (in addition to
the pp-wave spacetimes) also have this property.
Indeed, pp-waves provide exact solutions of string 
theory \cite{HS,amati},
and type-IIB superstrings in this background were shown 
to be exactly solvable
even in of the~presence of the~RR five-form field 
strength \cite{matsaev,mets}.
As a result the~spectrum of
the~theory can be explicitly obtained, and the number
of potential exact vacua for string theory is increased.
This work is expected to provide some hints for
the~study of superstrings on more general backgrounds.

\section{Analysis}

The classical equations of motion for a metric in string 
theory can be expressed in terms of $\sigma$-model 
perturbation theory \cite{cfmp}, through the Ricci tensor 
$R_{\mu \nu}$ and higher order corrections in
powers of the string tension scale $\alpha'$ and terms 
constructed from derivatives and higher powers of
the Riemann curvature tensor [e.g., $\frac{1}{2} \alpha^\prime
R_{\mu \rho \sigma \lambda} R_\nu \; ^{\rho \sigma \lambda}$].
It has long been known that vacuum pp-wave spacetimes are exact
solutions to string theory to all orders in $\alpha'$, 
and this was explicitly
generalized to non-vacuum (null radition) pp-wave solutions 
in \cite{HS}. The proof that all of the VSI spacetimes \cite{cppm} are  classical
solutions of the string equations
to all orders in $\sigma$-model 
perturbation theory \cite{amati} consists of showing that 
all higher order correction terms vanish \cite{amati},
and this follows immediately from the results of \cite{cppm}.
This can be demonstrated explicitly by direct calculation for the type III
example (\ref{example}) (see also \cite{bipo}).
It is perhaps surprising
that such a wide class of  VSI spacetimes, which contain 
a number of arbitrary functions,
have this property.

A more geometrical derivation 
of this result follows from 
the fact that the only non-zero
symmetric second-rank tensor covariantly constructed
from scalar invariants and
polynomials in the~curvature and their covariant derivatives
in VSI spacetimes is the Ricci tensor (which is proportional to $l_\mu
l_\nu$
for PP-type  O  spacetimes),
and hence
all higher-order terms in
the~string equations of motion automatically vanish \cite{HS}. 
More importantly, it
is possible to generalize this approach to include other bosonic massless
fields of the string.   For example, we can include a dilaton
$\Phi$
and an antisymmetric (massless field)
$H_{\mu v \rho}$. 
Let us assume that for VSI spacetimes 
\begin{eqnarray}
 \Phi & = & \Phi (u, \zeta, \ol{\zeta}), \label{phi} \\
H_{\mu \nu \rho} & = & A_{ij} (\mu, \zeta, \ol{\zeta}) 
\ell_{[\mu} \nabla_\nu x^i \nabla_{\rho]} x^j.\label{mu}
\end{eqnarray}
The field equations \cite{cfmp}
\begin{eqnarray}
2(\nabla \Phi)^2 -  \nabla^2 \Phi -  \frac{1}{12} H^2 = 0, \label{fe1} \\
\nabla_\lambda  H^{\lambda}_{\mu\nu}  - 2 (\nabla_\lambda \Phi)  H^{\lambda}_{\mu\nu} = 0, \label{fe2}
\end{eqnarray}
are then 
satisfied automatically
to leading order in $\sigma$-model 
perturbation theory (i.e., to order $\alpha^\prime$). This is 
clearly evident for VSI spacetimes with $\tau =0$
and with $\Phi=\Phi(u)$ and $A_{ij}=A_{ij}(u)$
and follows from the fact that $\ell$ is recurrent; this can be shown explicitly
by direct calculation for the type III
example (\ref{example}) and is known to be true for the pp-waves.
In these
spacetimes $H^2 = \nabla^2 \Phi = (\nabla \Phi)^2 =0$, and the only
non-trivial field equation is then \cite{cfmp}
$$ R_{\mu \nu} - \frac{1}{4} H_{\mu \rho \sigma}H_\nu \; ^{\rho \sigma}
- 2 \nabla_\mu \nabla_\nu \Phi =0.  $$
For PP-type $0$ spacetimes ($R_{\mu \nu} \propto \ell_\mu \ell_\nu)$
this equation has only one non-trivial component which then 
constitutes a single differential equation for 
the functions $\Phi$, $H_{\mu \nu \rho}$ and the metric functions
(e.g., for pp-wave spacetimes $\partial^2  H+ \frac{1}{18} A_{ij} A^{ij}
+ 2 \Phi^{\prime\prime} =0$ \cite{HS}).
Solutions with more general forms for $\Phi, H_{\mu \nu \rho}$ (than equations 
(\ref{phi}), (\ref{mu})) are possible for $\tau=0$ (and $\tau \neq 0$
spacetimes)
and the field equations reduce to an 
underdetermined set of differential equations with many arbitrary functions.

We can consider 
higher order corrections in $\sigma$-model
perturbation 
theory, which are of the form of second rank tensors and scalars
constructed from 
$\nabla_\mu \Phi$, $H_{\mu \nu \rho}$, the metric
and their derivatives.
Since at most two derivatives of $\Phi$ can appear in any second
rank tensor and given the form of the Riemann tensor,
all terms constructed from more that two $H_{\mu \nu \rho}$'s
and their derivatives and at least one Riemann tensor and one or more 
$\nabla_\mu \Phi$'s 
or $H_{\mu \nu \rho}$'s must vanish.  Also, for 
appropriately chosen $\Phi$ and $H_{\mu \nu \rho}$, all 
terms of the form $(\nabla \ldots \nabla H)^2$ 
must vanish.  
This can again be shown by direct calculation for the type III
example (\ref{example}) and was proven for the pp-waves in \cite{HS}.
Therefore, there are solutions to string 
theory to all orders in $\sigma$-model perturbation theory.
In addition, there are
clearly a large number of arbitrary functions (and many more than in the pp-wave case) 
in this class of solutions still to be determined. 

\section{Discussion}

It has been noted that massive fields can also be included
since their loop contributions can always
be expanded in powers of derivatives (the result will again
be polynomials in curvature which will vanish).
In addition, it has been shown \cite{HS} that exact pp-waves are exact solutions to string 
theory, even non-perturbatively.   It is therefore plausible that a wide  class of VSI solutions, which 
depends on a number of
arbitrary functions, 
are exact solutions to string theory 
non-perturbatively and worthy of further investigation.
In particular, the
singularity structure of the VSI string theory spacetimes can
be studied as in \cite{HS}.

Solutions of classical field equations for
which the~counter terms required to regularize
quantum fluctuations
vanish are also of importance because
they offer insights into the~behaviour of the~full quantum theory.
The~coefficients of quantum corrections
to Ricci flat solutions
of Einstein's theory of gravity in four dimensions
have been calculated up to two loops. In particular,
a class of Ricci flat (vacuum) Lorentzian 4-metrics, 
which includes the~pp-wave spacetimes
and some special Petrov type III or N spacetimes,
have vanishing counter terms up to and including two loops \cite{bipo}.
Thus these Lorentzian metrics suffer no quantum corrections
to all loop orders \cite{GG}.
In view of the~vanishing of all
quantum corrections in these spacetimes, it is possible that the VSI
metrics are of importance and merit further investigation.

{\em Acknowledgements}. I would like to thank 
R. Milson, V. Pravda, A. Pravdov\' a and R. Zalaletdinov for 
discussions, and V. Pravda and  A. Pravdov\' a for  further helpful comments on the manuscript.
This work was supported, in
  part, by a research grant from NSERC.

{}


\begin{thebibliography}{}

\bibitem{jordan}
P. Jordan, J. Ehlers and W. Kundt,  
Abh. Akad. Wiss. Mainz, Math.-Nat. no 2,
77 (1960).

\bibitem{deser} S. Deser, J. Phys. {\bf A8}, 1972 (1975).


\bibitem{gibb} G. W. Gibbons, Comm. Math. Phys.  {\bf 45}, 191 (1975).

\bibitem{amati} 
D. Amati and C. Klimcik,
Phys. Lett. {\bf B219} 443 (1989);
{\em ibid.} {\bf B210} 92 (1988).

\bibitem{HS} 
G.T. Horowitz and A.R. Steif, Phys. Rev. Lett.
{\bf 64} (1990) 260; {\em ibid.} Phys. Rev.
{\bf D42} 1950 (1990).

\bibitem{cppm}  V. Pravda, A. Pravdov\' a,  
A. Coley and R. Milson, Class. Quantum Grav. (gr-qc/0209024).


\bibitem{kramer}
D. Kramer, H. Stephani, M. MacCallum  and E. Herlt, 1980  
(Cambridge: Cambridge University Press)

\bibitem{kundt}
W. Kundt,  
Z.~Phys. {\bf 163} 77 (1961).

\bibitem{joly}
G.C. Joly, M.A.H. MacCallum,
Class. Quantum Grav. {\bf 7} 541 (1990).

\bibitem{energy}
C. Barcelo and M. Visser, gr-qc/0205066


\bibitem{matsaev} 
R.R. Metsaev, Nucl. Phys. {\bf B625} 
70 (2002). 

\bibitem{mets}
R.R.~Metsaev and A.~A.~Tseytlin, hep-th/0202109.

\bibitem{cfmp} C.G. Callan, D. Friedman, E.J. Martinec and 
M.J. Perry, Nucl. Phys. {\bf B262} 593 (1985).

\bibitem{bipo}
V. Pravda, Class. Quantum Grav.
{\bf 16} 3321 (1999).

\bibitem{GG} 
G.W. Gibbons, Class. Quantum Grav. {\bf 16} L71 (1999).



\end{thebibliography}
\end{document}